# Diffusion propagator metrics are biased when simultaneous multi-slice acceleration is used

L. Tugan Muftuler[1], Andrew S. Nencka[2,3], Kevin M. Koch[2,3]

[1]Department of Neurosurgery, [2]Department of Radiology, [3]Center for Imaging Research
Medical College of Wisconsin, 8701 Watertown Plk Rd. Milwaukee, WI, 53226, USA

Corresponding author:

L. Tugan Muftuler, PhD
Associate Professor of Neurosurgery
Medical College of Wisconsin.
8701 Watertown Plank Road
Milwaukee, WI 53226, USA
Email: lmuftuler@mcw.edu
Phone: +1(414) 955-7627

## ABSTRACT

Advanced diffusion MRI models are being explored to study the complex microstructure of the brain with higher accuracy. However, these techniques require long acquisition times. Simultaneous multi-slice (SMS) accelerates data acquisition by exciting multiple image slices simultaneously and separating the overlapping slices using a mathematical model. However, this slice separation is not exact and leads to crosstalk between simultaneously excited slices. Although this residual leakage is small, it affects quantitative MRI techniques such as diffusion imaging.

In this study, the effects of SMS acceleration on the accuracy of propagator metrics obtained from the MAP-MRI technique was investigated. Ten healthy volunteers were scanned with SMS accelerated multi-shell diffusion MRI acquisitions. Group analyses were performed to study brain regions typically affected by SMS acceleration. In addition, diffusion metrics from atlas based fiber tracts of interest were analyzed to investigate how propagator metrics in major fiber tracts were biased by 2- and 3-band SMS acceleration.

Both zero-displacement metrics and non-Gaussianity metrics were significantly altered when SMS acceleration was used. MAP-MRI metrics calculated from SMS-3 showed significant differences with respect to SMS-2. Furthermore, when shorter TR afforded by SMS acceleration was used, the characteristics of this bias have changed.

This has implications for studies using diffusion MRI with SMS acceleration to investigate the effects of a disease or injury on the brain tissues.

## 1. INTRODUCTION

MRI diffusion imaging has become a versatile tool to study brain tissues. While diffusion tensor imaging (DTI) is by far the most commonly used model to estimate bulk diffusion behavior in a voxel, simplified Gaussian diffusion approximation used in the model compromises its accuracy. Advanced diffusion MRI models are being explored to study the complex microstructure of the brain with higher accuracy. However, these techniques typically require long acquisition times to obtain sufficient samples of the three-dimensional diffusion space.

Simultaneous Multi-Slice (SMS) technique gained attention recently to accelerate diffusion MRI acquisitions. SMS shortens data acquisition time by exciting multiple slices simultaneously and separating the aliased slices using a mathematical model [1–5]. Typically, the overlapping slices are separated by estimating a set of reconstruction kernels from a calibration data using a least squares estimation approach. As a result, slice unaliasing is not exact and leads to information leakage between simultaneously excited slices [1,6–9]. Although, this residual leakage is small, it might affect quantitative MRI techniques such as diffusion imaging. In a diffusion image series, this persistent contamination of a voxel by the residues from the overlapping ones leads to a bias in estimating diffusion model parameters.

The effects of imperfect voxel unaliasing was investigated for functional MRI with in-plane GRAPPA acceleration [10] and SMS acceleration [11,12]. They noted that the residual leakage from imperfect unaliasing induced false correlations in voxel time courses, leading to compromised sensitivity and specificity in resting state connectivity analysis. Since diffusion MRI is essentially acquiring time course data while applying diffusion weighting gradients, the nature of these problems is similar in both techniques. However, the impact of SMS on diffusion MRI strongly depends on the relative direction and density of fiber tracts in overlapping voxels.

The impact of SMS slice leakage on diffusion tensor imaging (DTI) metrics has been previously studied in brain white matter by the Human Connectome Project group [13]. Using a conventional DTI acquisition with a single b-value of 1000s/mm$^2$ and 64 directions, they did not observe noticeable changes in fractional anisotropy (FA) in the major white matter tracts but reported a decrease in the estimated volume of the corpus callosum in tractography. This might be expected because DTI is more robust to noise and other artifacts compared to higher order diffusion MRI models [14]. Small changes in the actual diffusion orientation distribution function (ODF) due to SMS slice leakage may not translate to changes on a similar scale in parameter estimates for the diffusion tensor ellipsoid. On the other hand, more sophisticated diffusion MRI models that aim to fit the ODF more accurately might be sensitive to such biases. As a result, an advanced diffusion model might experience greater inaccuracy in the presence of slice leakage compared to DTI.

The effect of SMS on a multi-compartment diffusion model called Neurite Orientation Dispersion and Density Imaging (NODDI) [15] was recently reported by Bouyagoub et al [16]. The investigators acquired data with conventional single-band and SMS accelerations of 2 and 3. They reported poor agreement of Intraclass Correlation Coefficient (ICC) between NODDI parameter maps derived from single-band and SMS accelerations. The SNR was also shown to be spatially variable and generally lower in SMS acquisitions compared to the single-band.

The goal of the study presented here was to explore the effect of SMS further on a model-free advanced diffusion technique called Mean Apparent Propagator MRI (MAP-MRI). This is a mathematical framework that relates the diffusion MRI samples in q-space with the molecular displacements in tissues [17]. The diffusion signal attenuation in q-space, $E(\vec{q})$, is represented using a set of basis functions that are the eigenfunctions of the quantum-mechanical analog of the simple harmonic oscillator Hamiltonian. This framework makes it easy to derive a diverse set of rotation-invariant scalar parameters from the $E(\vec{q})$ that are sensitive to different features of the tissue microstructure. The first set of these scalar parameters is called zero-displacement metrics, which include return-to-origin probability (RTOP), return-to-axis probability (RTAP), return to plane probability (RTPP) and mean squared displacement (MSD). MAP-MRI also calculates a measure of non-Gaussianity (NG) in axial and radial directions. Initial studies comparing DTI and MAP-MRI demonstrated that MAP-MRI had much better sensitivity and accuracy in differentiating tissues with distinct structural features or anomalies [18]. However, the sensitivity of such advanced diffusion techniques could be compromised when SMS acceleration is used. Therefore, the experimenter needs to ensure that SMS does not confound the outcome of an experiment. This could be particularly critical for group analysis studies because the effects sought by the experiment could be suppressed or false positives might be induced by the slice leakage effects.

In this study, multi-shell diffusion data were acquired from volunteers with two-band and three-band SMS acceleration and with long and short TR values. A *Single-Group, Three Measurements ("Tripled T-Test")* model was implemented in General Linear Model (GLM) and group analyses were performed on each MAP-MRI parameter map to investigate the extent and severity of biases introduced by SMS acceleration and the impact of shorter TR facilitated by SMS acceleration.

## 2. METHODS

### 2.1. *Data Acquisition*:

This study was approved by the Institutional Review Board and prospective written consents were obtained from all participants. Data were acquired from ten healthy volunteers with no known neurological disorders (Age = 28±7.6 years, 6 females, 4 males). Experiments were conducted on a GE Healthcare Signa Premier 3.0T system (GE Healthcare, Waukesha WI), using a head coil array with 32-channels (Nova Medical, Wilmington MA). Vendor provided pulse sequence and image reconstruction software were used for the MAP-MRI acquisitions with 2-band and 3-band SMS accelerations. The pulse sequence was a single-shot Spin-Echo EPI and blipped-CAIPI scheme [19] with vendor-configured field-of-view shifts for optimal SNR. SMS reconstruction was performed using the product software, which is a hybrid-space coil-by-coil slice unaliasing approach. The calculation of reconstruction kernels for in-plane acceleration for this hybrid space approach was published earlier [20]. SMS reconstruction uses a similar formalism, calculating kernels from a single-band reference scan.

MAP-MRI data were collected using six b-values uniformly distributed between 1000s/mm$^2$ and 6000s/mm$^2$ with 96 directions. The q-space sampling was optimized for accuracy of MAP-MRI metrics using a genetic search algorithm published earlier [21]. From each participant, three SMS scans were acquired with acceleration factors of two (SMS-2) and three (SMS-3). The parameters for the first two scans were TE=90ms and TR=4900ms and 2mm isotropic resolution. The TR was kept long to allow for T1 relaxation and minimize spin history effects. Since the main advantage of SMS is to shorten TR and reduce total imaging time, an additional SMS-3 was acquired with the minimum allowed TR of 2650ms to test the effect of T1 relaxation on SMS. All other imaging parameters were matched between these scans. No in-plane acceleration was used.

### 2.2. *Preprocessing of diffusion image series*

Preprocessing of the diffusion data included distortion correction using the *topup* tool and motion and eddy current related distortion corrections using the *eddy* tool of the FSL software [22]. The three diffusion image series from a subject were not registered to each other before preprocessing or model fitting to minimize potential biases on q-space measurements that might be introduced by registration and interpolation errors. Each image set was visually inspected before and after the preprocessing to ensure that the image quality was not compromised by head motion or inaccurate distortion corrections.

### 2.3. *MAP-MRI model fitting*

MAP-MRI model fitting was performed using the DIPY Python software package [23]. The MAP-MRI model was fit with a radial order of six and positivity constraint [17] and Laplacian regularization [24].

This follows the recommendation of the developers of the DIPY MAP-MRI software that suggested using a low Laplacian weight of 0.05 together with the positivity constraint to guarantee non-negative propagator and reduce spurious oscillations in reconstruction. Additionally, the conventional DTI tensor fitting was also implemented to obtain FA maps, which were used primarily for the registration to standard space FA map (FMRIB58_FA.nii.gz). FSL's DTIFIT tool was used with weighted least squares option for this step. Full multi-shell data was used for the DTI fitting.

### 2.4. *Registration to standard space and statistical analysis*:

In order to perform group analysis, the diffusion maps from each subject needs to be registered to a standard space template. Typically, the first step is to correct for head motion between scans for each subject and then apply one transformation to standard space for all images from a subject. For the first step, the FA maps from SMS-3 long TR and SMS-3 short TR were first registered to the FA map from SMS-2 for each subject. FSL's FLIRT tool with 6-parameter affine transformation was applied for this step [22]. Then the respective affine transformations were applied to the remaining MAP-MRI metric images. For the second step, the FA map from the SMS-2 scan was registered to the standard space FA map (FMRIB58_FA_1mm.nii.gz image from FSL) using ANTS diffeomorphic registration [25]. This transformation was then applied to all the remaining registered FA and MAP-MRI maps from a subject to transform them to the standard space. This approach should preserve within-subject data consistency better compared to each scan being independently ANTS-transformed to the standard space. All registered images were visually inspected to check for artifacts or inaccurate registration.

Tract-Based Spatial statistics (TBSS) approach [26] was used for the analysis of group differences in major white matter tracts. However, the image registration step in the original TBSS pipeline with linear (FLIRT) and non-linear (FNIRT) registration was not used. Instead, the diffeomorphic ANTS registration was used as described above. Following the image registrations to the standard space, a white matter skeleton was generated for the group by running *tbss_skeleton* script applied on the ANTS registered FA images. An FA white matter skeleton image is generated by projecting the local FA values onto the skeleton. A threshold of FA > 0.2 was applied to include all major WM tracts and exclude superficial white matter and gray matter. Then, the same process was applied to the $RTOP^{1/3}$, RTAP, RTPP, MSD, NG mean, NG axial and NG radial images.

Voxelwise statistical tests were performed on the skeletonized FA and MAP-MRI images using FSL's *randomise* function [27,28], which employs a non-parametric approach to evaluate the General Linear Model (GLM) using a large number of random permutations to identify statistically significant effects. Randomise was run with 2000 permutations and Threshold-Free Cluster Enhancement (TFCE) option [29]. Each statistical test was corrected for multiple comparisons using a family-wise error (FWE) rate and results were thresholded using $p<0.05$. The design matrix and contrasts were generated for *Single-Group, Three Measurements ("Tripled T-Test")* following the formalism described in FSL GLM wiki page [30]. This model allows using three measures per subject and perform pairwise comparisons between SMS-2, SMS-3 long TR and SMS-3 short TR. Exchangability group values for randomise were set accordingly as described in FSL's GLM wiki page [30].

### 2.5. ***White matter atlas-based region of interest (ROI) analysis***:

The standard space white matter fiber tract atlas developed by John Hopkins University and available in FSL (JHU-ICBM-labels-1mm.nii.gz) was used to define different white matter fiber tracts and calculate average diffusion metrics in each of the fiber tracts. It is anticipated that different regions of the brain will experience different levels and directions of bias depending on the characteristics of diffusion in overlapping voxels. Eight major fiber tracts from the atlas were selected and MAP-MRI metrics were averaged in each fiber tract for each SMS acceleration. The selected fibers were genu, splenium and body of corpus callosum, superior longitudinal fasciculus (SLF), cingulum in the cingulate cortex, superior fronto-occipital (SFO) and posterior thalamic radiations (PTR). For FA, posterior limb of internal capsule (PLIC) showed significant bias with SMS-3 short TR; therefore, that fiber tract was included in FA plots. For SLF, cingulum, PTR and PLIC, right and left sides were studied separately as they might experience different levels and directions of bias depending on the voxels aliased onto each one. These were for illustration of the magnitude, direction and variance of bias with SMS acceleration. Statistical analyses were not performed in these individual fiber tracts since TBSS analysis provides that information on a voxel-by-voxel basis.

Fiber tract ROI plots were generated only for $RTOP^{1/3}$, NG-radial and FA to illustrate typical bias in selected fiber tracts. Illustrating similar plots for all MAP-MRI metrics was avoided since that would create an exhaustive set of plots. For the same reason, only a subset of the 68 fiber tracts listed in the atlas was selected as an example in order to demonstrate effects of SMS on most commonly studied fiber tracts. This analysis can be easily extended to other fiber tracts and other metrics if needed.

## 3. RESULTS

SMS affected most of the MAP-MRI metrics and the Fractional Anisotropy derived from the conventional DTI model. The effects were stronger with SMS-3 compared to SMS-2. Moreover, SMS acceleration of 3 with short and long TR values also showed a difference. All results reported here are with $p<0.05$, FWE corrected.

### 3.1. *Statistical Parametric Maps (TBSS analysis)*:

#### 3.1.1. SMS-3 versus SMS-2 (matching TR=4.9s):

Fig.1 illustrates TBSS analysis results comparing SMS-2 and SMS-3 with long TR values (4.9s). All three zero-displacement maps were significantly affected by the SMS-3 acceleration. There were widespread differences between $RTOP^{1/3}$, $RTAP^{1/2}$ and RTPP maps calculated from SMS-3 and SMS-2. $RTOP^{1/3}$, $RTAP^{1/2}$ and RTPP estimates were higher in the highlighted regions when SMS-3 was used. All three non-Gaussianity maps also had significant changes with SMS-3 acceleration. NG axial and NG mean values calculated from SMS-3 data were higher compared to those from the SMS-2. On the other hand, the effects were in the opposite direction for NG-radial, MSD and FA, where their values decreased with SMS-3 compared to SMS-2.

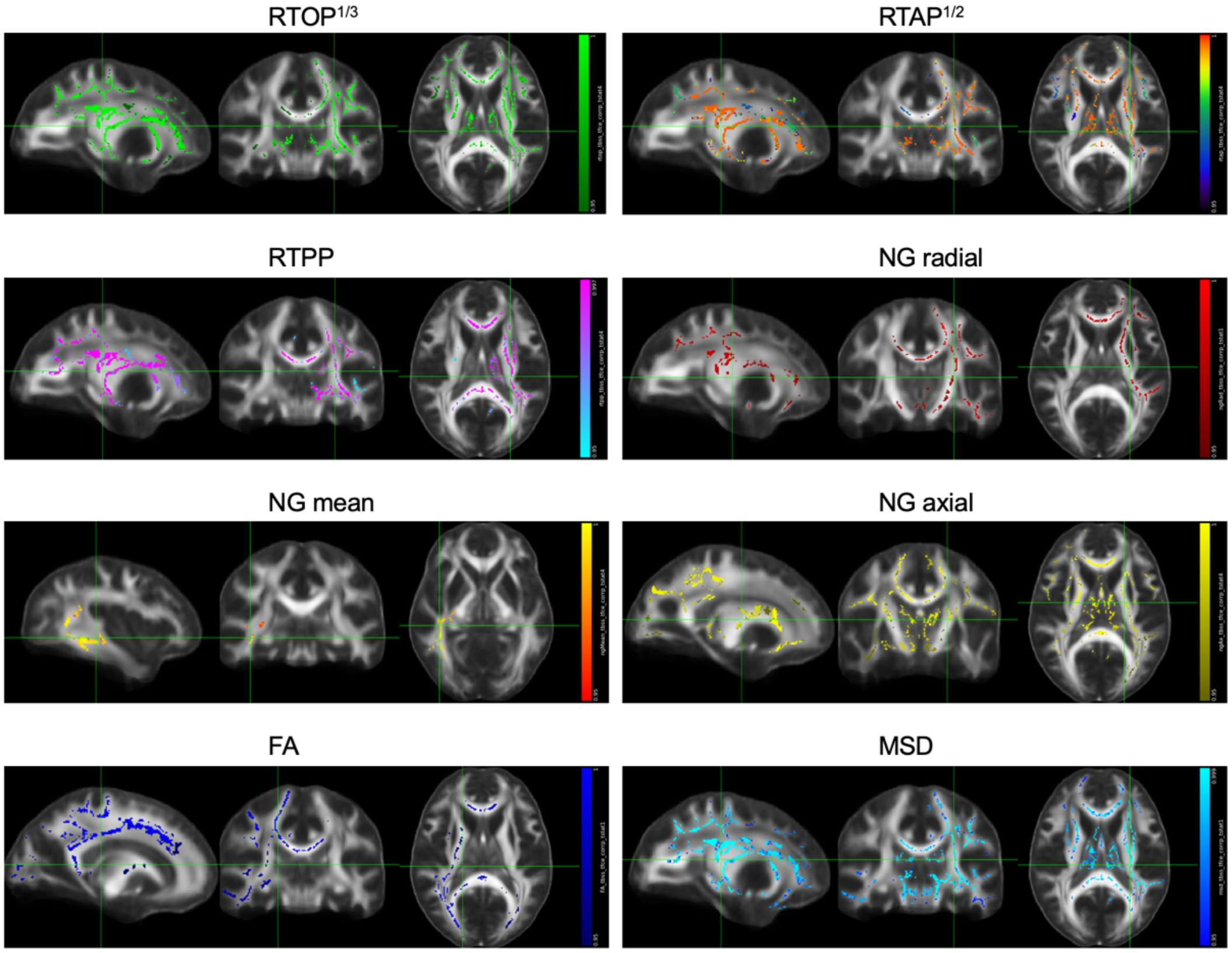

*Fig. 1. TBSS group analysis results comparing MAP-MRI maps from SMS-3 (TR = 4.9 s) and SMS-2 (TR = 4.9 s) acquisitions. Voxels with significant group differences are overlaid in color on group FA map. Images are displayed in the radiological convention.*

#### 3.1.2. SMS-3 (TR=2.65) versus SMS-2 (TR=4.9s):

An interesting finding was the effect of shorter TR afforded by the SMS acceleration. For SMS factor of 3, when the TR was reduced from 4.9s to 2.65s (minimum TR allowed by the SMS-3 acceleration), the resulting NG axial, NG mean and FA maps still showed significant differences with respect to the SMS-2 (Fig.2). FA had widespread decrease with SMS-3 (short TR) compared to SMS-2. This was in accord with the results with SMS-3 long TR, but the effects were more widespread with short TR. Similarly, the two non-Gaussianity metrics were higher with SMS-3 (short TR) compared to SMS-2, which were also in agreement with SMS-3 (long TR) results. The other metrics did not show significant differences.

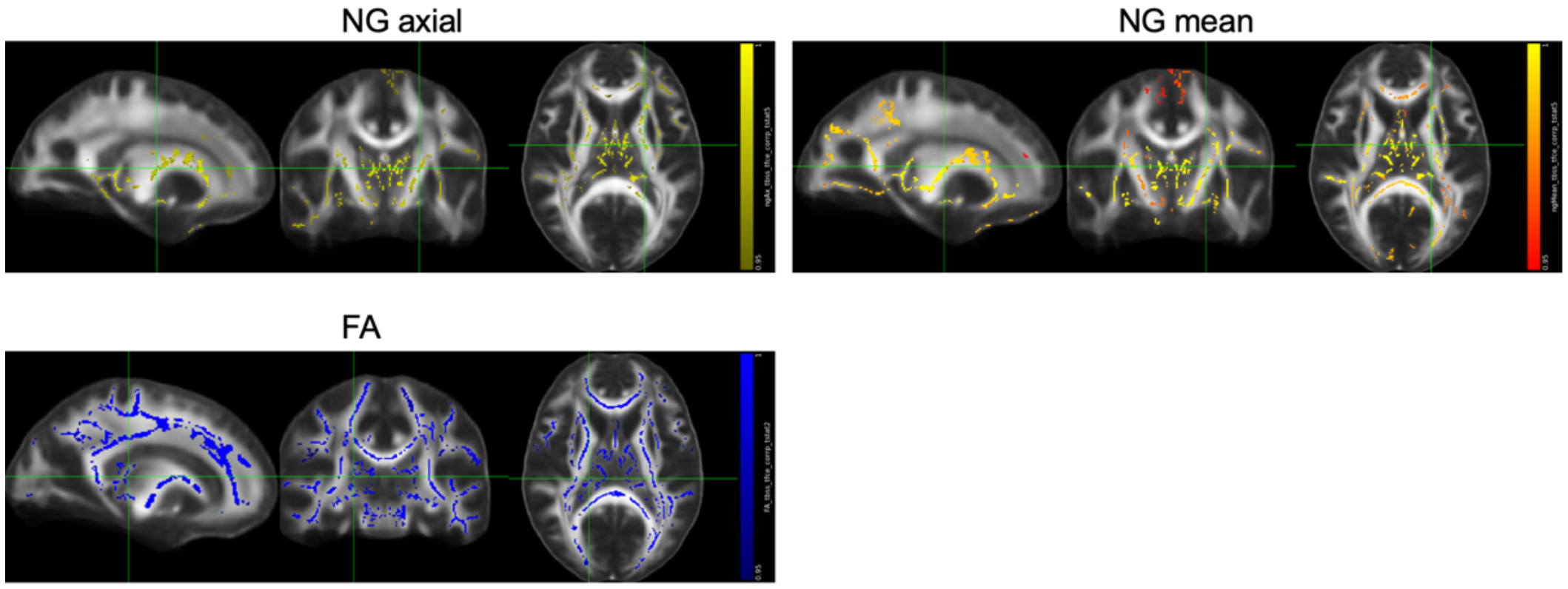

*Fig. 2. TBSS group analysis results comparing MAP-MRI maps from SMS-3 (TR = 2.65 s) and SMS-2 (TR = 4.9 s) acquisitions. Voxels with significant group differences are overlaid in color on group FA map. Images are displayed in the radiological convention.*

### 3.1.3. SMS-3 (TR=2.65) versus SMS-3 (TR=4.9s):

A comparison between SMS-3 with long and short TR (Fig.3) resembled patterns seen in Fig.1, which compared SMS-2 and SMS-3 with long TR. Overall, the results indicated that the effects of increased SMS acceleration were reduced by the shorter TR.

### 3.2. *White matter atlas-based ROI analysis*:

Although group analysis with TBSS shows typical trends with SMS acceleration, not all brain regions demonstrated the same direction of bias. Several atlas-based white matter fiber tracts of interest were selected as described earlier and $RTOP^{1/3}$, NG radial and FA values were averaged inside each tract. Figures 4, 5 and 6 show results from the selected fiber tracts. $RTOP^{1/3}$ values from SMS-3 (TR=4.9s) demonstrated similar increases in genu, splenium, corpus callosum, both sides of cingulum, but showed different patterns on the right and left sides of the SLF and PTR (Fig.4). SMS-3 (TR=2.65s) results were generally closer to those of SMS-2, except for PTR and corpus callosum. Average FA values were decreased genu, splenium, corpus callosum, SLF (right side) and cingulum (right side) for SMS-3 with both long and short TR. The values further dropped with SMS-3 short TR. On the other hand, average FA values in SLF (left side) cingulum (left side) and PLIC (left side) did not change much with SMS-3 long TR but dropped significantly with SMS-3 short TR. NG radial generally decreased with SMS-3 (TR=4.9s) in most of the fiber tracts studied (Fig.5). It only increased in the PTR on the right side and did not change noticeably in the SLF-right. Similar to $RTOP^{1/3}$ results, SMS-3 (TR=2.65s) tracts were generally closer to those of SMS-2, except for PTR-left, to some extent in the corpus callosum, splenium and genu.

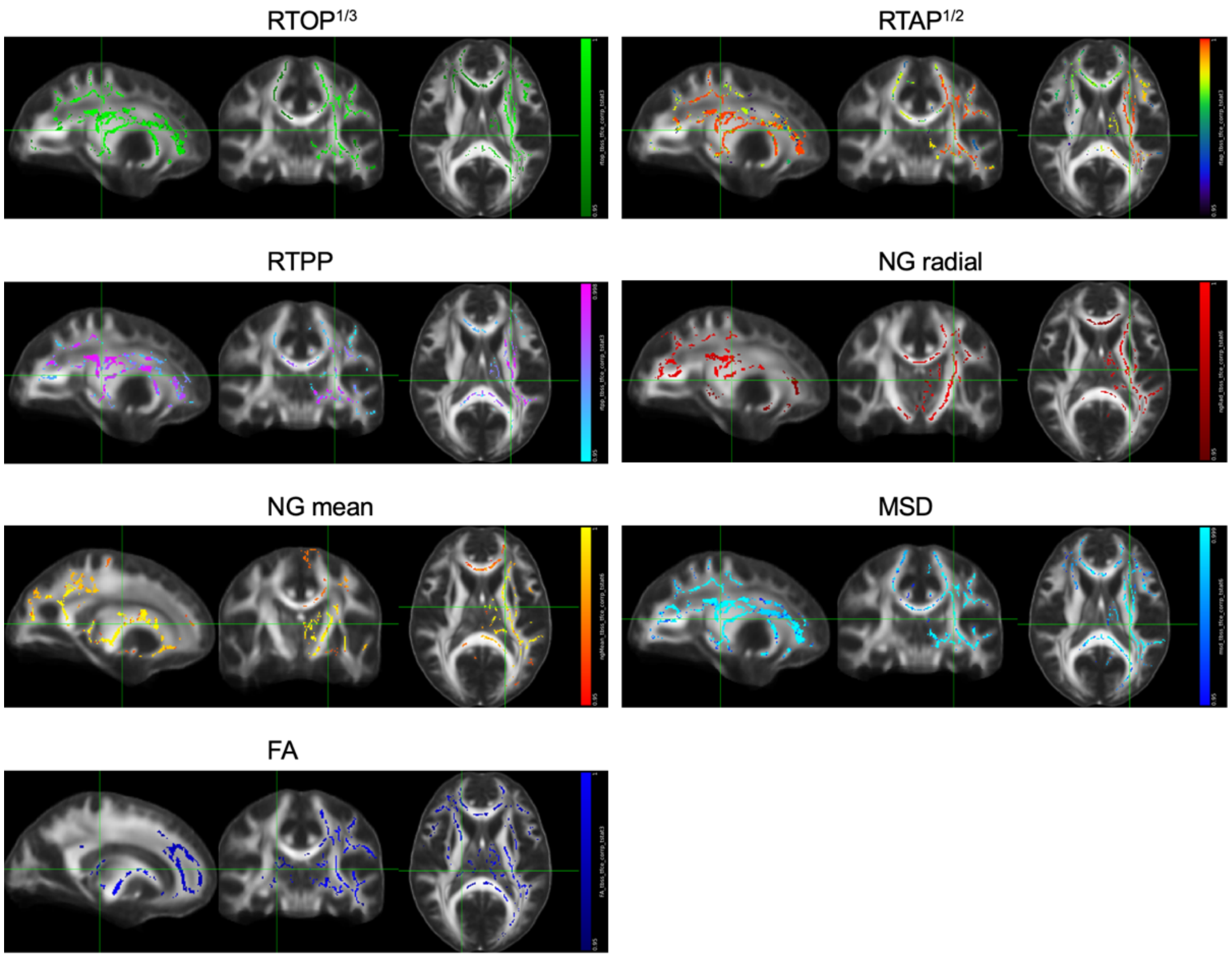

*Fig. 3. TBSS group analysis results comparing MAP-MRI maps from SMS-3 (TR = 4.9 s) and SMS-3 (TR = 2.65 s) acquisitions. Voxels with significant group differences are overlaid in color on group FA map. Images are displayed in the radiological convention.*

***Supplementary data:*** An additional MAP-MRI scan without SMS acceleration (single-band) was also acquired with scan parameters matching the SMS-2 and SMS-3 TR=4.9s scans. However, those were not used for the main analyses because the manufacturer informed us that when SMS is turned off, coil channels are combined using conventional sum-of-squares method. However, when SMS is turned on, the scanner software uses SENSE = 1 method. Since this might introduce a confounding effect [31,32], and our goal was to test vendor-provided pulse sequence and reconstruction, the single-band acquisitions were not used for the main results that were presented here. However, analyses that included single-band scans were included as supplementary data to provide some insights.

In the supplementary data, the comparison of FA and MAP-MRI metrics calculated from SMS-2 and single-band did not show statistically significant differences. This suggested that difference in coil channel combination was not the main factor driving the observed differences. However, the lack of control over the image reconstruction with the product software and the presence of a potential confounding effect required that single-band comparisons should only be provided as a supplementary information.

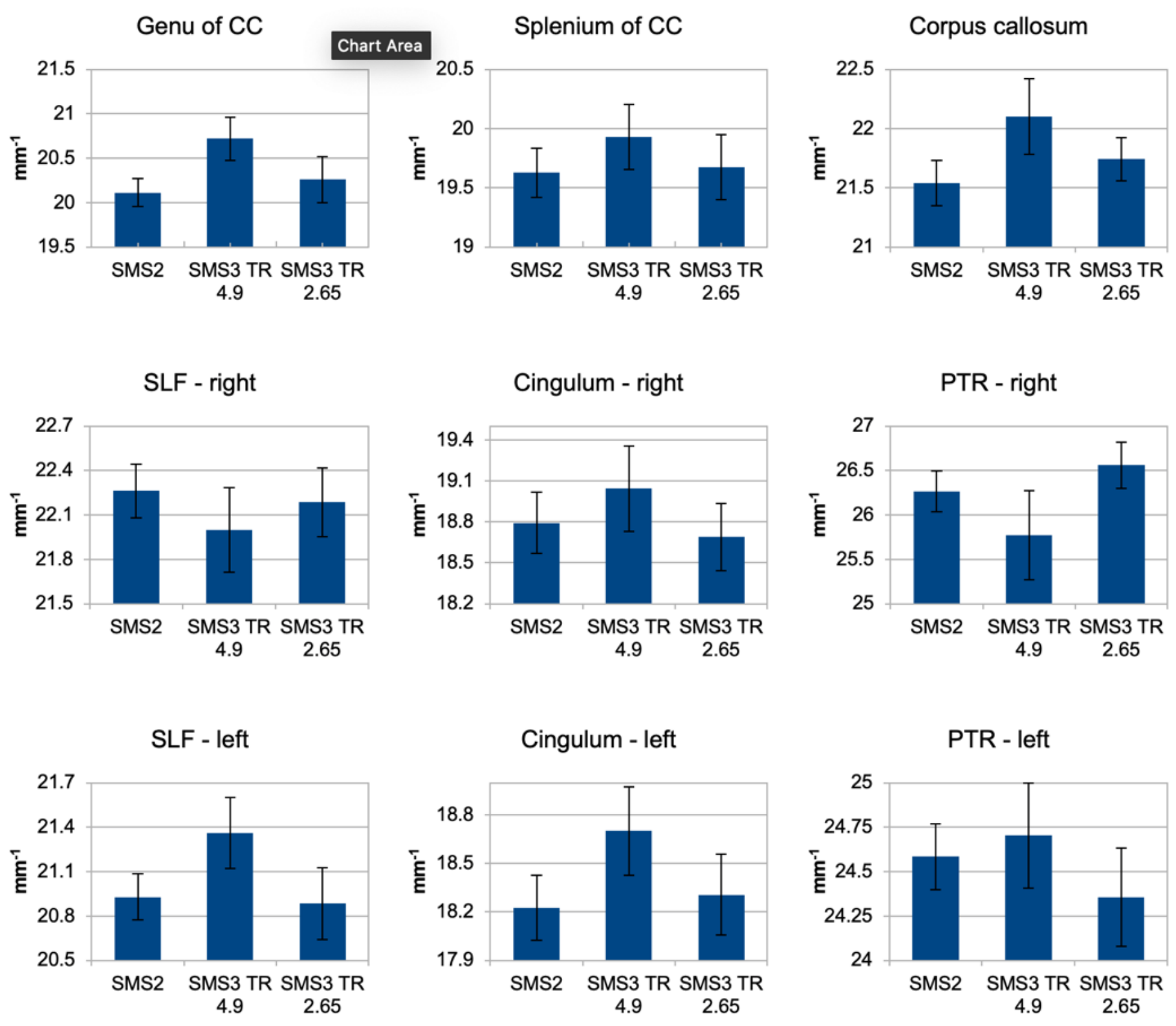

*Fig. 4. Average RTOP$^{1/3}$ plots in selected ROIs from JHU white matter atlas for different SMS accelerations. Error bars indicate standard error.*

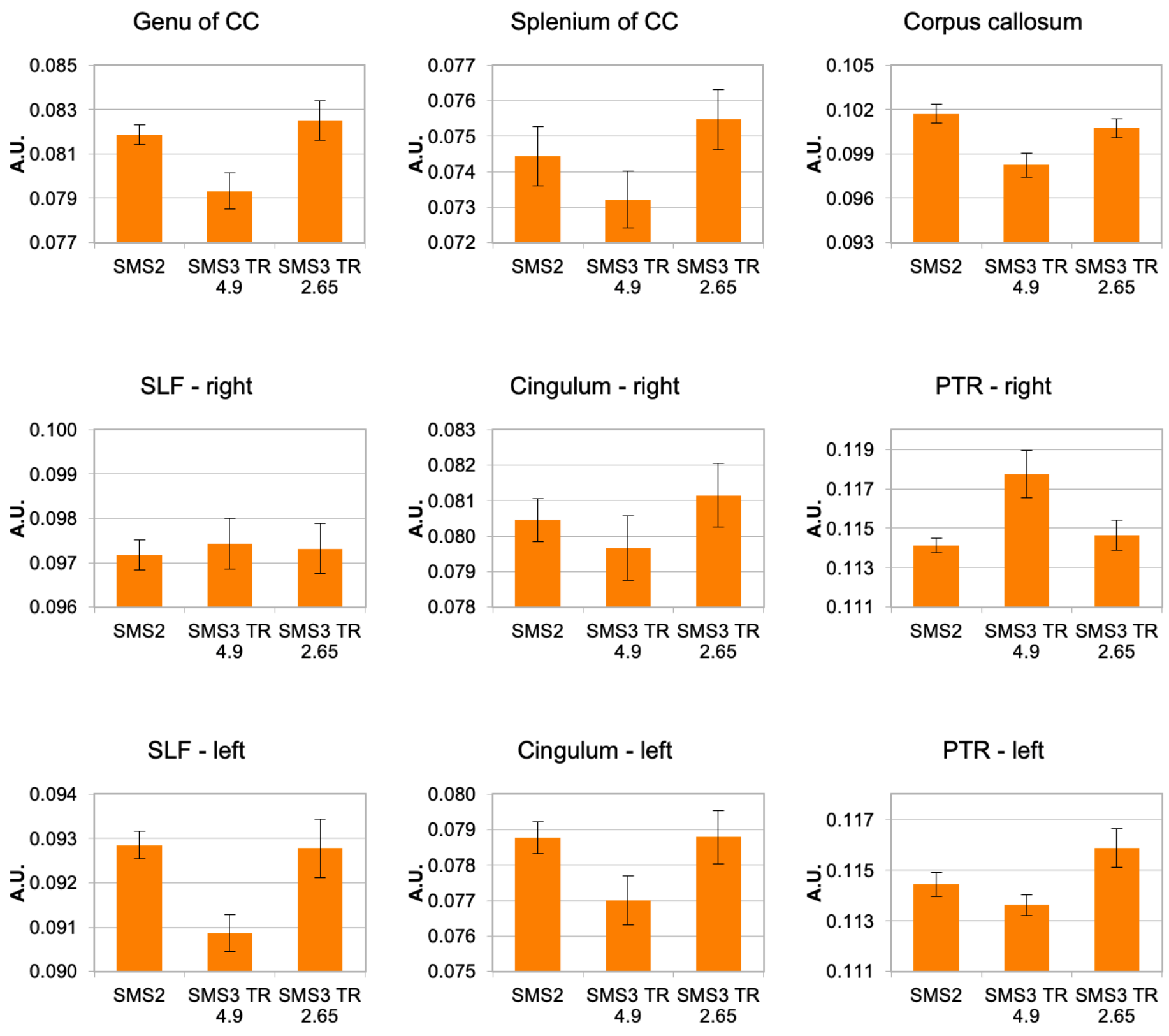

*Fig. 5. Average NG-radial plots in selected ROIs from JHU white matter atlas for different SMS accelerations. Error bars indicate standard error.*

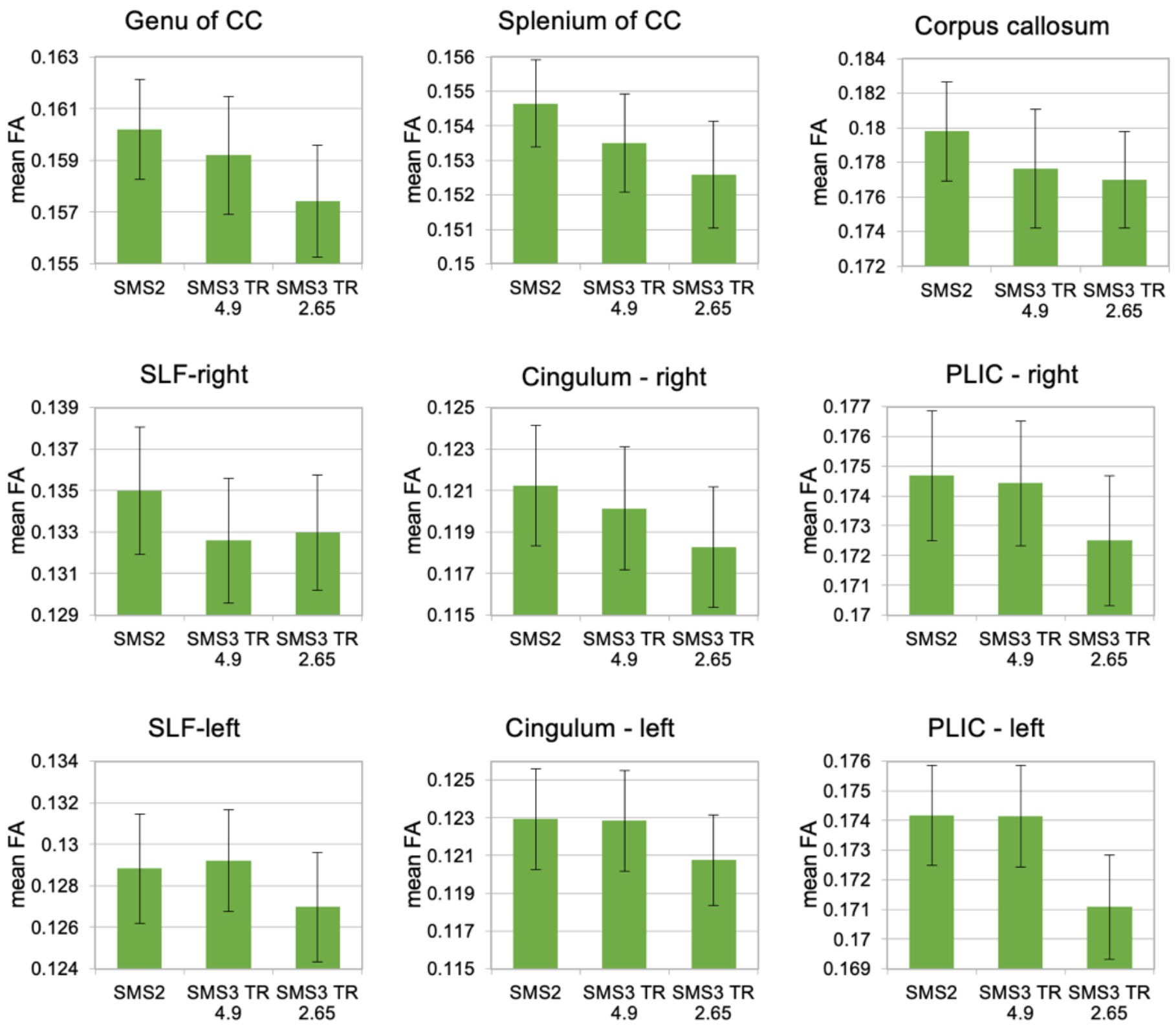

*Fig. 6. Average FA plots in selected ROIs from JHU white matter atlas for different SMS accelerations. Error bars indicate standard error.*

## 4. DISCUSSION

The results presented here show evidence of measurement bias introduced by SMS acceleration for diffusion MRI. The q-space measurements ($E(\vec{q})$) in a voxel in the brain white matter and various metrics estimated by the MAP-MRI technique and FA from DTI were significantly affected even with modest SMS accelerations used in these experiments. It appears that the signal leakage between aliased slices in SMS acceleration caused enough perturbation to bias the $E(\vec{q})$ measurements and resulting coefficient estimates for the MAP-MRI basis functions and derived metrics.

In the majority of the brain white matter, the RTOP estimates increased with increasing SMS acceleration. This is in accord with widespread decrease in MSD in the same white matter regions. Both results indicate that the estimated ODF size decreased due to slice leakage from aliased slices. This could happen, for instance, if a voxel with higher diffusion (hence reduced signal with a particular diffusion weighting gradient vector) is mixed with a voxel with lower diffusion (hence higher signal) with the same diffusion weighting. However, not all brain regions showed the same direction of bias as shown in figures 4 and 5. Some white matter tracts were biased in the opposite direction. This is expected since this bias will be dictated by the diffusion characteristics of the overlapping voxels. This is also in accord with an earlier study reporting spatially varying effects in NODDI model when SMS was used [16].

The results from RTPP and RTAP suggest that the ODF estimates were strongly affected both in the axial direction (biasing RTPP) and radial direction (biasing RTAP).

Furthermore, when shortest available TR was used that was afforded by SMS acceleration, the influence of SMS on MAP-MRI estimates changed. This is possibly driven by the T1 relaxation steady state effects (spin history). Interestingly, this effect counterbalanced the unwanted bias from aliased slices in some brain regions and some of the metrics that were significantly different with SMS-3 long TR (Fig.1) did not show the same effects here. NG axial, NG mean and FA were still affected by the increased SMS factor.

The consistency of bias in the fiber tracts that showed similar differences across subjects was an interesting finding. Since there is usually some variation in brain shapes and sizes across subjects, a voxel in a fiber tract in the primary slice might mix with different fiber tracts from the aliased slices in different subjects. However, this consistency seen here might be expected. As long as the aliased voxels contain fiber tracts with different directions with respect to the primary voxel, they will perturb the resulting signal attenuation distribution in q-space and resulting MAP-MRI metrics. An aliased fiber tract making a $40^{o}$ angle with respect to the fiber tract in the primary voxel might be biasing less than one making a $90^{o}$ angle, but they will perturb q-space attenuation measurements in a similar manner. Furthermore, the signal leakage into the primary voxel might be from large structures such as CSF or deep brain gray matter from the aliased slices. In those cases, the primary voxel might be aliased with the same structure in different subjects, even though it might be a different section of the same structure. For instance, ventricles span several slices. So, a fiber tract voxel contaminated by CSF in the ventricles in the aliased slice will show similar signal leakages across subjects despite some variations in brain sizes. In the end, our statistical analysis captures regions where the bias was sufficiently consistent across subjects compared to the variance.

An earlier study that investigated the effect of SMS on FA using a 3-band SMS acceleration and conventional DTI acquisition did not find a measurable effect [13], whereas our results clearly showed a bias on FA. This could be due to the fact that they only used a single low b-value of 1000 s/mm$^2$ and 64 directions compared to the multi-shell acquisition used here. An acquisition with multiple shells and high b-values will be tapping into diffusion at different scales, affecting the DTI fitting.

Our findings are in accord with a recent study that investigated effects of SMS on NODDI [16], where they reported significant changes in NODDI parameter maps with SMS acceleration using ICC analysis. Our study complements these earlier works and reports how SMS acceleration affects group analysis results. Unlike these earlier studies that used a specific diffusion model to represent diffusion patterns in a voxel (ellipsoid for DTI, or multi-compartment diffusion for NODDI), MAP-MRI does not use an microstructural model to represent diffusion patterns. Instead, it uses the 3-dimensional distribution of signal attenuation in q-space itself for a given acquisition and derives metrics from those measurements. As long as sufficiently high b-values and directional sampling were used in the acquisition and MAP basis functions with order 6 (or higher) are used, the representation of ODF should more accurately reflect the actual diffusion patterns inside a voxel compared to such model-based diffusion techniques. While MAP-MRI is reported to be more sensitive to subtle changes in tissue morphology both in brain white and gray matters [18], this higher sensitivity could be compromised by SMS acceleration.

In this study, the experiments were designed to minimize confounding effects. All sequence parameters were matched between acquisitions, except for the SMS value (and TR for the additional acquisition). However, the study had some limitations. First of all, comparisons with single-band were not included in the main analysis due to the potential confounding effect caused by differences in product image reconstruction. In order to understand the effects of SMS acceleration on diffusion MRI microstructure analysis, future studies should include single-band as reference and use in-house reconstruction to ensure consistent image reconstruction across all acquisitions. Future studies should also include repeated acquisitions to test within-session and between session reproducibility.

## 5. CONCLUSION

Studies utilizing SMS acceleration for advanced diffusion MRI acquisitions need to investigate the impact of SMS acceleration on the calculated diffusion parameters. Our earlier analysis showed that the bias introduced by SMS is relatively small in FA and MD maps if single-shell diffusion acquisition and simple DTI model were used [33]. This is possibly because of the simple diffusion model of DTI and its muted sensitivity to complex diffusion patterns. However, more advanced diffusion techniques such as MAP-MRI have higher sensitivity to subtle contamination from imperfect slice separation in SMS reconstruction. Therefore, accuracy of diffusion data might be compromised due to slice leakage introduced by SMS.

**Acknowledgements**

Funding: This study is supported in part by Daniel M. Soref charitable foundation and GE Healthcare technology development grant.

## Supplementary data with single-band acquisitions

Additional comparisons between single-band acquisitions and both of the SMS-3 acquisitions are provided here to give readers further insights. All pulse sequence parameters of the single-band matched the SMS-2 and SMS-3 (TR=4.9s), except for the SMS factor. FA and MAP-MRI metrics calculated from SMS-2 did not show statistically significant differences in TBSS analysis. However, some plots from selected white matter tracts show notable changes even with SMS-2.

It should be noted that the differences in image reconstruction (coil channel combination) between SMS scans and single-band might have contributed to some of the differences here. Although lack of significant differences between SMS-2 and single-band suggest that it was not the main factor, the results presented here should be viewed with the understanding that changes in reconstruction might still amplify or dampen the effects of SMS acceleration.

***Fig. S1. SMS-3 (TR = 4.9 s) versus single-band (TR = 4.9 s)***

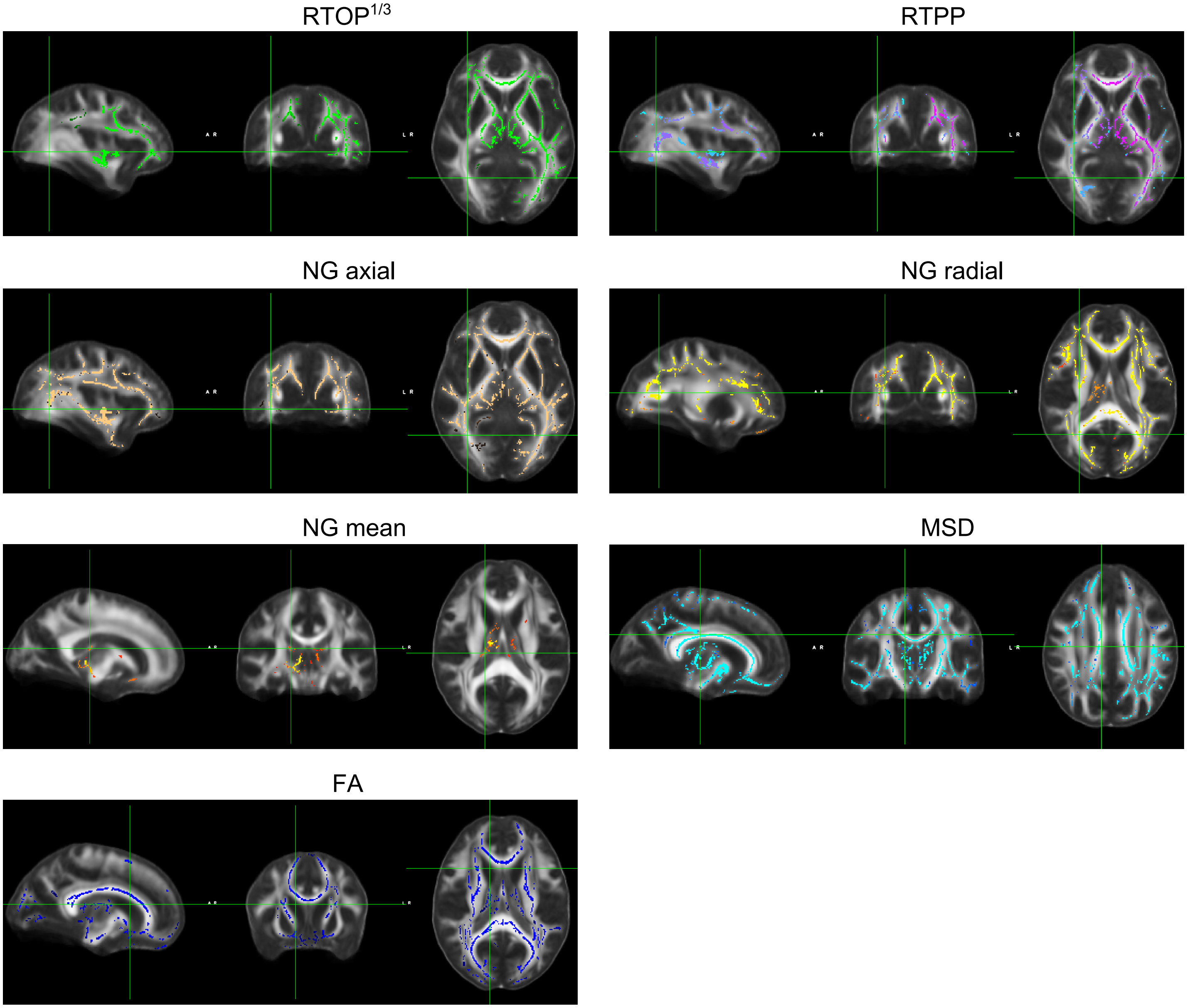

***Fig. S2. SMS-3 (TR = 2.65 s) versus single-band (TR = 4.9 s)***

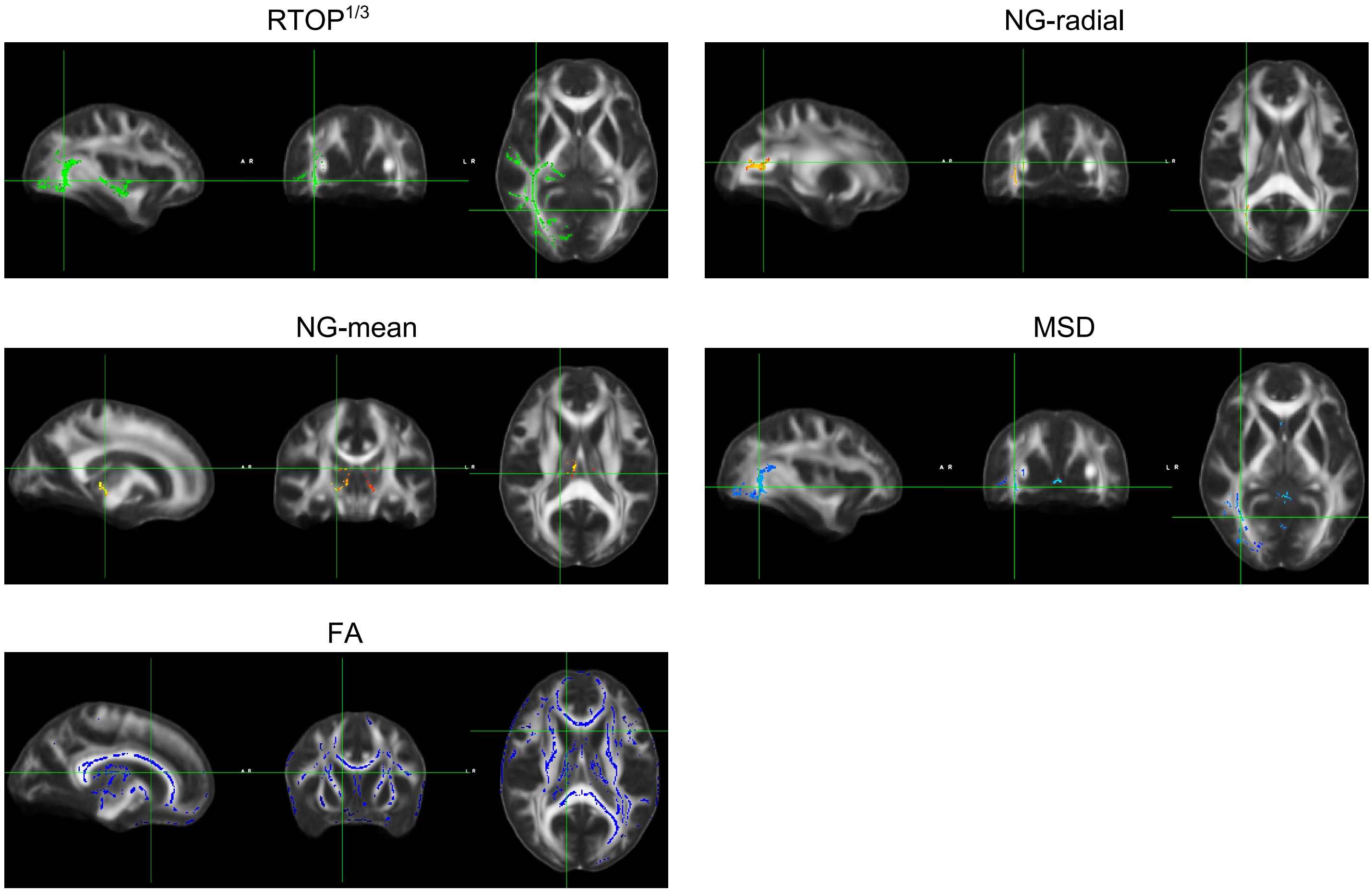

***Fig. S3. RTOP$^{1/3}$ plots in selected ROIs from JHU white matter atlas***

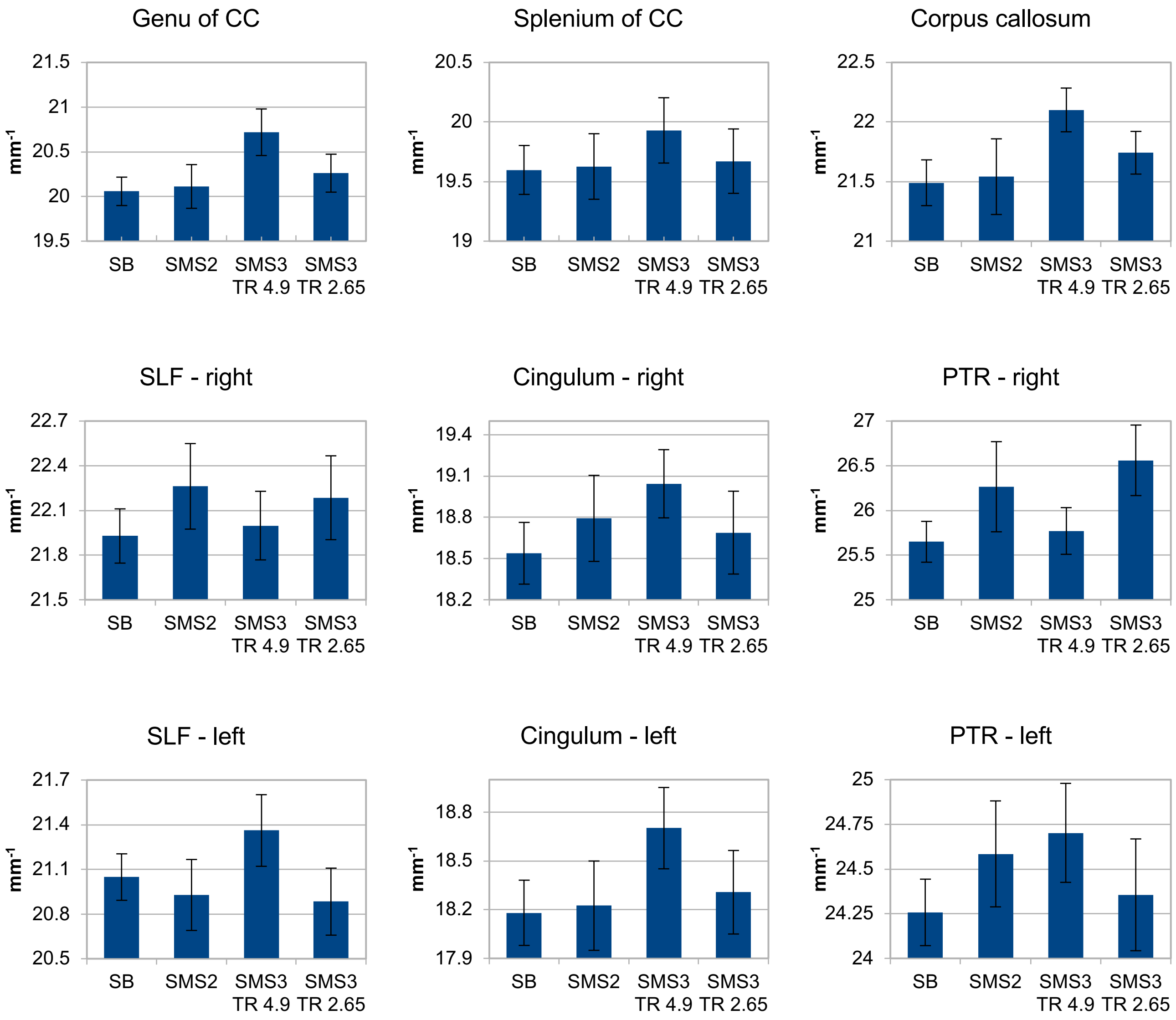

***Figure S4. NG-radial plots for selected ROIs from JHU white matter atlas***

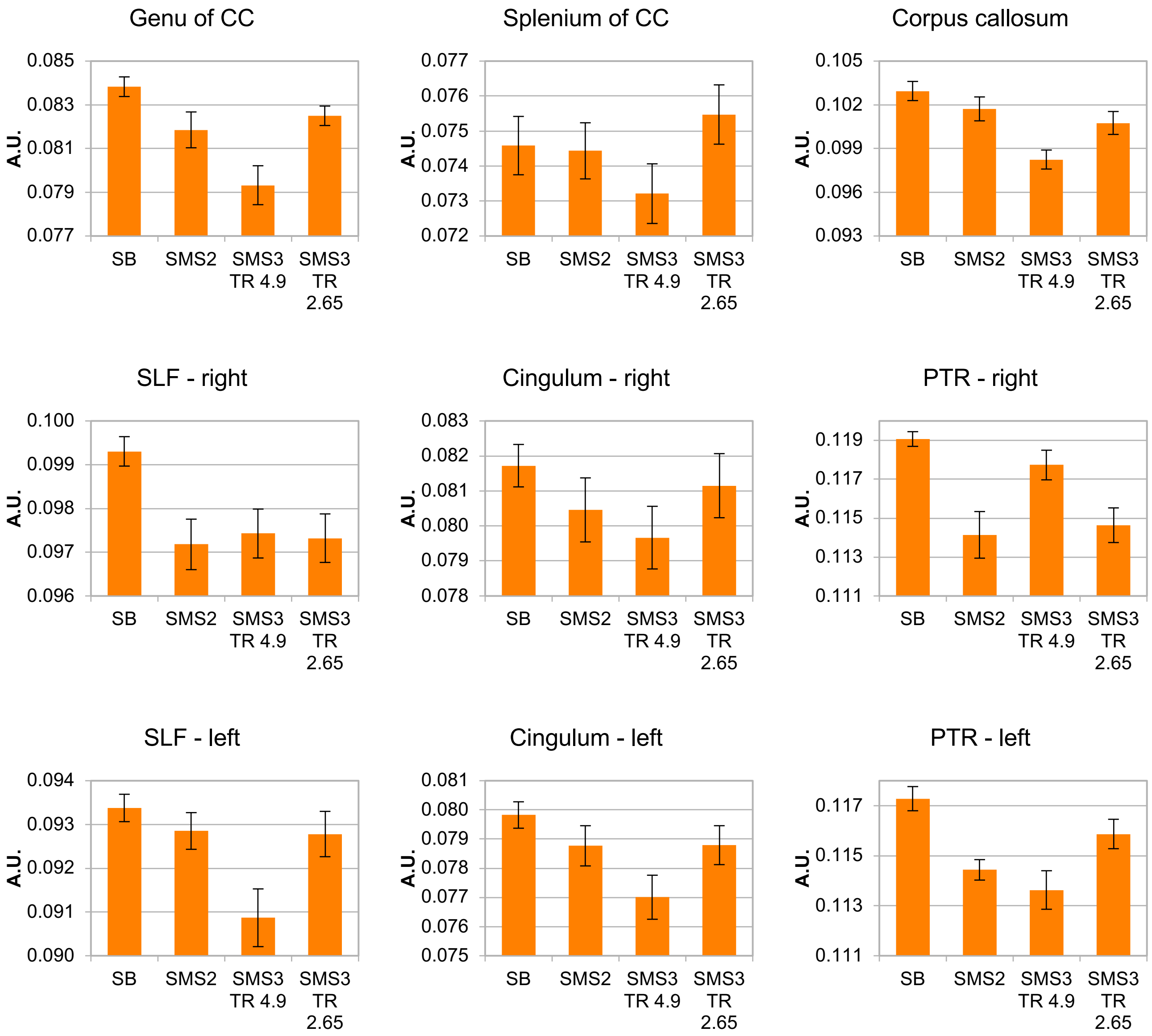